\documentclass[aps,prl,color,pdflatex,citeautoscript,superscriptaddress,twocolumn]{revtex4-1}
\usepackage[T1]{fontenc}
\usepackage{graphicx}
\usepackage{dcolumn}
\usepackage{amsmath,amssymb,bbold,bm}
\usepackage{hyperref}
\hypersetup{colorlinks=true,citecolor=blue,linkcolor=blue,urlcolor=blue,pdfstartview=FitH,pdfpagemode=PageWidth}
\usepackage{amsfonts}
\usepackage{listings}
\usepackage{braket}
\usepackage{float,latexsym}
\usepackage{multirow}
\usepackage{tikz}

\usepackage{color}
\definecolor{dkgreen}{rgb}{0,0.5,0}

\newcommand{\comment}[1]{}{}
\begin{document}

\title{Approximating the Sachdev-Ye-Kitaev model with Majorana wires}
\author{Aaron Chew}
\affiliation{Department of Physics and Institute for Quantum Information and Matter, California Institute of Technology, Pasadena, CA 91125, USA}
\author{Andrew Essin}
\affiliation{Department of Physics, University of California, Davis, CA 95616, USA}
\author{Jason Alicea}
\affiliation{Department of Physics and Institute for Quantum Information and Matter, California Institute of Technology, Pasadena, CA 91125, USA}
\affiliation{Walter Burke Institute for Theoretical Physics, California Institute of Technology, Pasadena, CA 91125, USA}

\date{\today}

\begin{abstract}
  The Sachdev-Ye-Kitaev (SYK) model describes a collection of randomly interacting Majorana fermions that exhibits profound connections to quantum chaos and black holes.  We propose a solid-state implementation based on a quantum dot coupled to an array of topological superconducting wires hosting Majorana zero modes.  Interactions and disorder intrinsic to the dot mediate the desired random Majorana couplings, while an approximate symmetry suppresses additional unwanted terms.  We use random matrix theory and numerics to show that our setup emulates the SYK model (up to corrections that we quantify) and discuss experimental signatures.
  
\end{abstract}

\maketitle

%%%%%%%%%%%%%%%%%%%%%%%%%%%%%%%%%%%%%%%%%%%%%%%%

{\bf \emph{Introduction.}}~Majorana fermions provide building blocks for many novel phenomena. 
As one notable example, Majorana-fermion zero modes \cite{ReadGreen, 1DwiresKitaev} capture the essence of non-Abelian statistics and topological quantum computation \cite{kitaev,TQCreview}, and correspondingly now form the centerpiece of a vibrant experimental effort \cite{mourik12,das12,finck12,Churchill,DeFranceschi,Nadj-Perge,Kurter,Xu,AlbrechtExponential,Gul2,DengPhaseTransition,Chains2}.  More recently, randomly interacting Majorana fermions governed by the `Sachdev-Ye-Kitaev (SYK) model' \cite{Sachdev,KitaevTalks,Maldacena} were shown to exhibit sharp connections to chaos, quantum-information scrambling, and black holes---naturally igniting broad interdisciplinary activity (see, e.g., \cite{Polchinski,YouSYK, FuSYK, Danshita, Jevicki, Gross, Banerjee, WittenSYK, GurauSYK, ColterSYK, DavisonSYK, JensenSYK, BiSYK, Gu, MarcelSYK}).  The goal of this paper is to exploit hardware components of a Majorana-based topological quantum computer for a tabletop implementation of the SYK model, thus uniting these very different topics.

The SYK Hamiltonian reads
\begin{equation}
  H_{\rm SYK} = \sum_{1\leq i<j<k<l\leq N} J_{ijkl} \gamma_i \gamma_j \gamma_k \gamma_l,
  \label{SYKmodel}
\end{equation}
where $\gamma_{i = 1,\ldots,N}$ denote Majorana fermions with `all-to-all', Gaussian-distributed random couplings $J_{ijkl}$ satisfying
\begin{equation}
    \langle J_{ijkl} \rangle = 0,~~\langle J_{ijkl} J_{i'j'k'l'} \rangle = \delta_{i,i'}\delta_{j,j'}\delta_{k,k'}\delta_{l,l'} \frac{3! \bar J^2}{N^3}.
  \label{J_properties}
\end{equation}
At large $N$ the model is solvable and exhibits rich behavior.  Most remarkably, for temperatures satisfying $\bar{J}/N \ll T \ll \bar J$ the SYK model enjoys approximate conformal symmetry and, similar to black holes, is maximally chaotic as diagnosed by out-of-time-ordered correlators.  These properties are expected for a holographic dual to quantum gravity, and there has been much interest in the corresponding bulk theory \cite{JensenSYK,Gross2}. 

\begin{figure}
	\includegraphics[width=8.5cm]{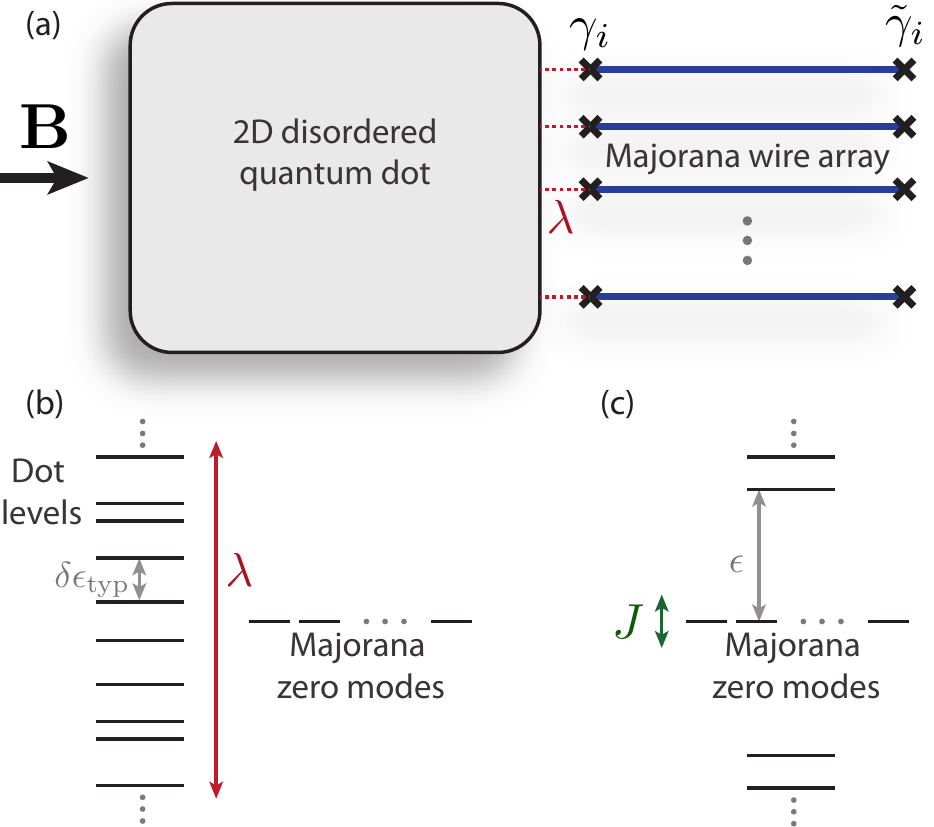}
	\caption{(a) Device that approximates the SYK model using topological wires interfaced with a 2D quantum dot.  The dot mediates disorder and four-fermion interactions among Majorana modes $\gamma_{1,\ldots,N}$ inherited from the wires, while Majorana bilinears are suppressed by an approximate time-reversal symmetry.  (b) Energy levels pre-hybridization.  The dot-Majorana hybridization energy $\lambda$ is large compared to $N \delta \epsilon_{\rm typ}$, where $N$ is the number of Majorana modes and $\delta \epsilon_{\rm typ}$ is the typical dot level spacing; this maximizes leakage into the dot.  (c) Energy levels post-hybridization.  The $N$ absorbed Majorana modes enhance the energy $\epsilon$ to the next excited dot state via level repulsion; four-Majorana interactions occur on a scale $J < \epsilon$. }
	\label{Setup}
\end{figure}

Laboratory realizations of Eq.~\eqref{SYKmodel} face intertwined hurdles: First, hybridizing Majorana fermions naively yields bilinears of the form $iM_{jk} \gamma_j \gamma_k$ as the dominant couplings, yet these are absent from the Hamiltonian.  Second, generating all-to-all couplings requires abandoning locality for the Majorana fermions.  And finally, the host platform must carry sufficient disorder to at least approximate independence among the large number of random $J_{ijkl}$'s.  References~\onlinecite{Danshita,MarcelSYK} proposed SYK-model platforms using cold atoms and topological insulators, respectively.  We instead envision a realization [Fig.~\ref{Setup}(a)] that exploits Majorana zero modes germinated in proximitized semiconductor nanowires \cite{1DwiresLutchyn,1DwiresOreg}---a leading experimental architecture for topological quantum information applications \cite{mourik12,das12,finck12,Churchill,DeFranceschi,Kurter,AlbrechtExponential,Gul2,DengPhaseTransition}.  

More precisely, we explore an array of such wires interfaced with a disordered quantum dot that mediates coupling among the constituent Majorana modes and randomizes the corresponding zero-mode wavefunctions.  Unwanted Majorana bilinears are suppressed by an approximate time-reversal symmetry \cite{TewariInvariant} that, importantly, is preserved by the dominant sources of disorder expected in the dot.  Interactions intrinsic to the dot instead generate the desired all-to-all four-Majorana couplings, thus approximating the SYK model up to corrections that we quantify (and which appear generic for any physical realization).  
We discuss several future directions that our approach spotlights, including tunneling experiments that provide a natural first probe of SYK physics.  

{\bf \emph{Setup.}}~We begin with the Hamiltonian for a clean, single-subband proximitized wire:
\begin{eqnarray}
  H_{\rm wire} &=& \int_x \bigg{[}\psi^\dagger\left(-\frac{\partial_x^2}{2m}-\mu -h \sigma^x -i \alpha \sigma^y \partial_x\right)\psi 
  \nonumber \\
  &+& \Delta(\psi_\uparrow \psi_\downarrow + H.c.) + \cdots\bigg{]},
  \label{Hwire}
\end{eqnarray}
which features Zeeman coupling $h$ generated by a magnetic field ${\bf B}$, spin-orbit coupling $\alpha$, and proximity-induced pairing $\Delta$.  Together these ingredients allow the formation of Majorana zero modes $\gamma, \tilde \gamma$ at the wire ends over a chemical potential window centered around $\mu = 0$ \cite{1DwiresLutchyn,1DwiresOreg}.  Crucially, the terms explicitly displayed above respect a time-reversal transformation $\mathcal{T}$ that sends $\psi \rightarrow \psi$, $i\rightarrow -i$ and thus satisfies $\mathcal{T}^2 = +1$ \cite{TewariInvariant}.  Additional couplings denoted by the ellipsis can in general violate $\mathcal{T}$ since it is not a true microscopic symmetry.  Nevertheless, we will assume that such perturbations are negligible, which is not unreasonable at low densities appropriate for the topological regime.  (See Discussion for further comments.) 
Under the approximate $\mathcal{T}$ symmetry the Majorana-zero-mode operators transform as $\gamma \rightarrow \gamma$ and $\tilde \gamma \rightarrow -\tilde \gamma$.  The opposite signs acquired by $\gamma, \tilde \gamma$ ensure that $\mathcal{T}$ commutes with the ground-state fermion parity $P = i\gamma \tilde \gamma$, as it must.  

Consider now $N$ topological wires `plugged into' a 2D disordered quantum dot [Fig.~\ref{Setup}(a)], such that the Majoranas $\gamma_{1,\ldots,N}$ that are even under $\mathcal{T}$ hybridize with the dot while their partners $\tilde \gamma_{1,\ldots,N}$ decouple completely.  The full architecture continues to approximately preserve $\mathcal{T}$ provided $(i)$ the dot carries negligible spin-orbit coupling and $(ii)$ the ${\bf B}$ field orients in the plane of the dot so that orbital effects are absent.  Here the setup falls into class BDI, which in the free-fermion limit admits an integer topological invariant $\nu \in \mathbb{Z}$ \cite{KitaevClassification, RyuClassification} that counts the number of Majorana zero modes at each end; interactions collapse the classification to $\mathbb{Z}_8$ \cite{FidkowskiKitaev1,FidkowskiKitaev2}.  In essence our device leverages nanowires to construct a topological phase with a free-fermion invariant $\nu = N$: All bilinear couplings $iM_{jk} \gamma_j\gamma_k$ are forbidden by $\mathcal{T}$ and thus \emph{cannot} be generated by the dot under the conditions specified above.  We exploit the resulting $N$ Majorana zero modes to simulate SYK-model physics mediated by disorder and interactions native to the dot, similar in spirit to Refs.~\onlinecite{YouSYK,MarcelSYK}.

Figures \ref{Setup}(b) and (c) illustrate the relevant parameter regime.  The dot-Majorana hybridization energy $\lambda$ satisfies $\lambda \gg N \delta \epsilon_{\rm typ}$, where $\delta \epsilon_{\rm typ}$ denotes the typical dot level spacing.  This criterion enables the dot to absorb a substantial fraction of \emph{all} $N$ Majorana zero modes as shown below.  The dot's disordered environment then efficiently `scrambles' the zero-mode wavefunctions, though we assume that their localization length $\xi$ exceeds the dot size $L$.  More quantitatively, we take the mean-free path $\ell_{\rm mfp} \ll L$ to maximize randomness and the dimensionless conductance $g = k_F \ell_{\rm mfp} > 1$ such that $L < \xi$.  Turning on four-fermion interactions couples the disordered Majorana modes with typical $J_{ijkl}$'s that are smaller than the energy $\epsilon$ to the next excited state (which as we will see is enhanced by level repulsion compared to $\delta \epsilon_{\rm typ}$).  This separation of scales allows us to first analyze the disordered wavefunctions in the non-interacting limit and then explore interactions projected onto the zero-mode subspace.  We next carry out this program using random-matrix theory, which is expected to apply in the above regime \cite{BeenakkerTransport,Aleiner}.  

{\bf \emph{Random-matrix-theory analysis.}}~We model the dot as a 2D lattice composed of $N_{\rm dot} \gg N$ sites hosting fermions $c_{a = 1,\ldots,N_{\rm dot}}$ \footnote{We assume spinless fermions for simplicity; spin can be introduced trivially since we impose $\mathcal{T}^2 = 1$ symmetry.}.  In terms of physical dot parameters we have $N_{\rm dot} \sim (L/\ell_{\rm mfp})^2$.  The Hamiltonian governing the dot-Majorana system is $H = H_0 + H_{\rm int}$, with $H_0$ and $H_{\rm int}$ the free and interacting pieces, respectively.  We employ a Majorana basis and write $c_a = (\eta_a + i\tilde \eta_a)/2$, where $\eta_a$ is even under $\mathcal{T}$ while $\tilde \eta_a$ is odd (similarly to $\gamma_i, \tilde \gamma_i$).  In terms of 
\begin{eqnarray}
  \Gamma = [\eta_1\cdots\eta_{N_{\rm dot}}; \gamma_1 \cdots \gamma_N]^T,~~\tilde \Gamma = [\tilde \eta_1\cdots \tilde \eta_{N_{\rm dot}}]^T,
\end{eqnarray}
$H_0$ takes the form
\begin{equation}
  H_0 = \frac{i}{4}\left[ \begin{array}{ccc}
  \Gamma^T && \tilde \Gamma^T
  \end{array}
  \right] 
  \left[ \begin{array}{ccc}
  0 & M \\
  -M^T & 0
  \end{array}
  \right]
  \left[ \begin{array}{ccc}
  \Gamma \\
  \tilde \Gamma
  \end{array}
  \right].
  \label{H0}
\end{equation}
Time-reversal $\mathcal{T}$ fixes the zeros above but allows for a general real-valued $(N_{\rm dot} + N)\times N_{\rm dot}$-dimensional matrix $M$.  (The matrix is not square since we discarded the $\tilde \gamma_i$ modes that trivially decouple.)  One can perform a singular-value decomposition of $M$ by writing $\Gamma = \mathcal{O} \Gamma'$ and $\tilde \Gamma = \tilde{\mathcal{O}} \tilde \Gamma'$. 
Here $\mathcal{O},\tilde{\mathcal{O}}$ denote orthogonal matrices consisting of singular vectors, i.e., the matrix $\Lambda \equiv \mathcal{O}^T M \tilde{\mathcal{O}}$ only has non-zero entries along the diagonal.  
Writing $\Gamma' = [\eta_1'\cdots\eta'_{N_{\rm dot}}; \gamma_1' \cdots \gamma_N']^T$ and similarly for $\tilde \Gamma'$, the Hamiltonian becomes
\begin{equation}
  H_0 = \frac{i}{2} \sum_{a = 1}^{N_{\rm dot}} \epsilon_a \eta_a' \tilde \eta_a'
\end{equation}
where $\epsilon_a \equiv \Lambda_{aa}$ are the non-zero dot energies.  Most importantly, $\gamma'_{i = 1,\ldots,N}$ drop out and form the modified $N$ Majorana zero modes guaranteed by $\mathcal{T}$ symmetry.  

We are interested in statistical properties of the associated Majorana wavefunctions in the presence of strong randomness.  To make analytic progress we assume (for now) that all elements of $M$ in Eq.~\eqref{H0} are independent, Gaussian-distributed random variables with zero mean and the same variance, corresponding to the chiral orthogonal ensemble \cite{VerbaarschotRMT,StephanovRMT}. This form permits Cooper pairing of dot fermions---an inessential detail for our purposes---and also does not enforce the strong-hybridization criterion $\lambda \gg N \delta \epsilon_{\rm typ}$.  We will see that the Majorana wavefunctions nevertheless live almost entirely in the dot as appropriate for the latter regime.  

The probability density for such a random matrix $M$ is \cite{BeenakkerReview2} $P(M) \propto \exp\left[{-\frac{\pi^2}{8 N_{\rm dot} \delta\epsilon_{\rm typ}^2}\text{Tr}(M^T M)}\right]$.  
%\begin{equation}
%  P(M) \propto e^{-\frac{\pi^2}{8 N_{\rm dot} \delta\epsilon_{\rm typ}^2}\text{Tr}(M^T M)}.  
%\end{equation}
Because $P(M)$ is invariant under $M \rightarrow \mathcal{O}^T M \tilde{\mathcal{O}}$, the singular-vector matrices $\mathcal{O}, \tilde{\mathcal{O}}$ are uniformly distributed over the spaces $O(N_{\rm dot}+N)$ and $O(N_{\rm dot})$, respectively.  In particular, the Majorana wavefunctions $\phi_i$ corresponding to $\gamma'_{i}$ are the final $N$ columns of a random element of $O(N_{\rm dot}+N)$.  For large $N_{\rm dot}+N$ the distribution of wavefunction components is asymptotically Gaussian \cite{GuhrRMT,BeenakkerTransport}:
\begin{equation}
  \langle \phi_{i,I} \rangle = 0,~~~~\langle \phi_{i,I} \phi_{j,J}\rangle = \frac{\delta_{i,j}\delta_{I,J}}{N_{\rm dot} + N} \approx \frac{\delta_{i,j}\delta_{I,J}}{N_{\rm dot}}.
  \label{phi_properties}
\end{equation}
Summing $\phi_{i,I}^2$ over the dot sites thus gives unity up to corrections of order $N/N_{\rm dot}$, i.e., the dot swallows the Majorana modes as claimed.

Once absorbed by the dot, the $N$ Majorana zero modes repel the nearby energy levels.  Random matrix theory allows us to estimate the energy $\epsilon$ to the first excited dot state.  References~\cite{SilversteinWishart, BaiWishart} show that the smallest eigenvalue for the Wishart matrix $M^TM$ approaches $(\sqrt{a}-\sqrt{b})^2 v$, where $M$ is an $a\times b$ matrix with variance $v$ for each element.  
The energy $\epsilon$ is the square root of this eigenvalue.  For our matrix $M$ we thus obtain
\begin{equation}
  \epsilon \approx \frac{1}{\pi}N\delta\epsilon_{\rm typ}.
  \label{epsilon}
\end{equation}
The enhancement compared to $\delta\epsilon_{\rm typ}$ [sketched in Fig.~\ref{Setup}(c)] isolates the $N$ Majorana modes from adjacent levels, justifying projection onto the zero-energy subspace.  

Let us now examine a general $\mathcal{T}$-invariant four-fermion interaction among dot fermions, $H_{\rm int} = \sum_{abcd} U_{abcd} c_a^\dagger c_b^\dagger c_c c_d$.  Projection follows from $c_a \rightarrow \frac{1}{2} \sum_i \phi_{i,a} \gamma_i'$, 
which yields 
\begin{eqnarray}
  H &\rightarrow& \sum_{1 \leq i<j<k<l \leq N} J_{ijkl} \gamma_i' \gamma_j' \gamma_k' \gamma_l'
  \\
  J_{ijkl} &=& \frac{1}{2^4} \sum_{abcd} U_{abcd} \sum_{p} s_p \phi_{p(i) a} \phi_{p(j) b} \phi_{p(k)c} \phi_{p(l)d}.
  \label{Jijkl}
\end{eqnarray}
The $p$ sum runs over permutations of $ijkl$, and $s_p = \pm 1$ is the parity of permutation $p$.  Notice that only the part of $U_{abcd}$ that is asymmetric under swapping any pair of indices contributes to $J_{ijkl}$.  For density-density interactions including Coulomb---where $U_{abcd} \propto \delta_{ad}\delta_{bc}$---all $J_{ijkl}$ consequently vanish.  This in fact is a virtue that underlies compatibility of SYK physics with randomness in our setup.  Density-density interactions would project nontrivially only if potential disorder $\delta \mu_a c_a^\dagger c_a$ did as well, but the latter would generate unwanted Majorana bilinears that tend to spoil SYK properties.  Other physical couplings such as current-current interactions produce non-zero $J_{ijkl}$.  

Emulating the SYK model requires that the $J_{ijkl}$'s encode all-to-all Majorana interactions and form independent random variables whose correlations obey Wick's theorem.  
Using Eq.~\eqref{phi_properties} one reproduces Eq.~\eqref{J_properties} with 
\begin{equation}
  %\bar J^2 = \frac{4!4! N^3}{3! 2^8 N_{\rm dot}^4} \sum_{abcd} (U^{\rm as}_{abcd})^2 \sim 1/N_{\rm dot}^\alpha.
  \bar J^2 = \frac{3 N^3}{8 N_{\rm dot}^4} \sum_{abcd} (U^{\rm as}_{abcd})^2 \sim \frac{N^3}{N_{\rm dot}^\alpha}.
  \label{J_variance}
\end{equation}
Here $U^{\rm as}_{abcd}$ denotes the antisymmetric part of $U_{abcd}$. The exponent $\alpha$ on the right side is interaction-dependent.  An (unphysical) non-local interaction with $(U^{\rm as}_{abcd})^2 = {\rm constant}$ yields $\alpha = 0$, while a local $U^{\rm as}_{abcd}$ with support only for $bcd$ `near' $a$ instead yields $\alpha = 3$.  

\begin{figure*}
	\includegraphics[width=\textwidth]{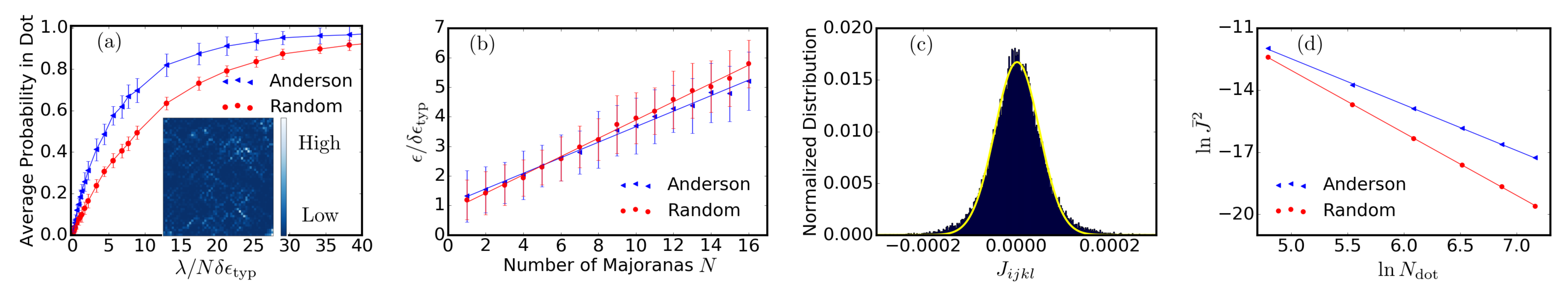}
	\caption{(a) Average absorption of Majorana wavefunctions into the dot versus the hybridization strength $\lambda$ with $N = 16$ zero modes.  Inset: probability density for a Majorana wavefunction swallowed and randomized by the dot of size $51 \times 51$.  (b) Enhanced level repulsion of the first excited dot state $\epsilon$ by $N$ absorbed Majorana modes; cf.~Figs. \ref{Setup}(b) and (c).  (c) Histogram of $J_{ijkl}$ couplings obtained from local current-current interactions on a dot of size $21 \times 21$, together with a Gaussian fit (solid line).  (d) Scaling of the variance $\propto \bar J^2$ of these couplings versus $N_{\rm dot}$. }
	\label{Numerics}
\end{figure*}

Equation~\eqref{J_variance} implies all-to-all coupling but does not guarantee independence of the $J_{ijkl}$'s.  Since there are ${N \choose 4} \sim N^4$ such couplings and $N \times N_{\rm dot}$ independent Majorana-wavefunction components in the dot, a necessary condition for the latter property is
\begin{equation}
  N_{\rm dot} \gtrsim N^3.
  \label{Ndot_relation}
\end{equation}
Corrections to Wick's theorem persist even in this regime, however.  For example, Eq.~\eqref{phi_properties} yields
\begin{eqnarray}
  \langle J_{ijkl}J_{klmn}J_{ijmn} \rangle &\propto& \frac{1}{N_{\rm dot}^6}\sum_{abcdef} U^{\rm as}_{abcd}U^{\rm as}_{cdef}U^{\rm as}_{abef} \sim \frac{1}{N_{\rm dot}^\beta},
  \nonumber \\ \label{WickViolation}
\end{eqnarray}
whereas in the SYK model such correlations vanish.  (Note that our system still preserves the statistical SO$(N)$ `flavor' symmetry corresponding to rotations among the Majorana fermions that is present in the SYK model.)  A local interaction implies $\beta = 5$; Eq.~\eqref{WickViolation} then decays faster with $N_{\rm dot}$ compared with $\langle J_{ijkl}^2\rangle^{3/2}$. In this sense the $J_{ijkl}$'s asymptotically form independent Gaussian random variables as in Ref.~\onlinecite{Danshita}.  Corrections to Wick's theorem do nevertheless introduce a proliferation of new Feynman diagrams that may qualitatively alter SYK-model physics over some energy scales.  The Appendix elaborates on this interesting issue.

{\bf \emph{Numerics.}}~We now semi-quantitatively validate random-matrix-theory predictions using a more physically motivated Hamiltonian.  Consider first the free part,
\begin{equation}
  H_0 = -\sum_{a \neq b}t_{ab}c_a^\dagger c_b + \sum_{a} V_a c_a^\dagger c_a  + \lambda \sum_{i = 1}^N\gamma_i (c_{a_i} - c_{a_i}^\dagger).
\end{equation}
Here $V_a$ is an uncorrelated Gaussian disorder landscape with zero mean and variance $\bar V^2$. In the $\lambda$ hybridization term, Majorana $\gamma_i$ couples to a single dot site $a_i$.  For the hoppings $t_{ab}$, we consider uniform nearest-neighbor tunnelings of strength $t$ (yielding an Anderson model) and compare results with purely random, arbitrary-range hopping satisfying $\langle t_{ab}\rangle = 0, \langle t_{ab} t_{a'b'} \rangle = t^2$ (yielding a random-matrix model).  All data below correspond to $\bar V = t$ with adjacent Majorana modes separated by two or three dot sites.   Unless specified otherwise $\lambda = t/2$, the dot system size is $31\times 31$, and results are disorder-averaged over many configurations [20 for Fig.~\ref{Numerics}(a), 50 for (b) and (d), and 500 for (c)].

Figure~\ref{Numerics}(a) corresponds to $N = 16$ and plots the fraction of the Majorana mode wavefunctions absorbed by the dot---averaged over all $16$ zero modes---versus $\lambda/(N\delta\epsilon_{\rm typ})$.  For both the Anderson and random-matrix models the fraction is of order one at $\lambda/(N\delta\epsilon_{\rm typ}) \gtrsim 4$, eventually saturating to unity as in random matrix theory.  The inset shows the probability density for a zero-mode wavefunction nearly fully absorbed by the dot, obtained from an $N = 1$ Anderson model; the wavefunction appears thoroughly randomized and loses all information about its original position (in this case, the center).  Figure~\ref{Numerics}(b) illustrates level repulsion of the excitation energy $\epsilon$ (normalized by the level spacing $\delta \epsilon_{\rm typ}$) versus $N$.  Note that the dot almost completely absorbs all zero modes up to the largest $N$ shown.  The random matrix model yields a slope that agrees within $\sim 5\%$ with Eq.~\eqref{epsilon} obtained from random matrix theory, while the Anderson model agrees within $\sim 20\%$.  

Next we include a local current-current interaction
\begin{equation}
  H_{\rm int} = U\sum_{\langle ab \rangle} c_a^\dagger \nabla c_a \cdot c_b^\dagger \nabla c_b,
\end{equation}
with $\nabla$ a lattice gradient, projected into the zero-mode subspace.  Figure~\ref{Numerics}(c) plots a histogram of the resulting $J_{ijkl}$ couplings (in units of $U$) using an Anderson model with $N = 8$ and a $21\times 21$ dot.  The data agrees well with a Gaussian distribution; see solid line.  Finally, Fig.~\ref{Numerics}(d) illustrates the $N_{\rm dot}$-dependence of the variance $\propto\bar J$ for $J_{ijkl}$ [recall Eq.~\eqref{J_properties}] with $N = 8$.  The Anderson model yields a scaling close $1/N_{\rm dot}^2$---slower than $1/N_{\rm dot}^3$ result from random matrix theory [Eq.~\eqref{J_variance}].  We attribute this difference primarily to localization effects that effectively reduce the system area.  As a check, the random-matrix model, which should not suffer localization due to the non-local hoppings, indeed yields the expected $1/N_{\rm dot}^3$ scaling.

{\bf \emph{Discussion.}}  We showed that in certain regimes our Majorana wire/quantum dot setup can emulate the SYK model up to very generic corrections.  Chiefly, we invoked an approximate time-reversal symmetry that suppresses bilinears, strong dot-Majorana coupling that delocalizes and randomizes the wavefunctions, level repulsion that suppresses pollution of the zero-mode subspace by additional dot levels, and sufficient randomness to approximate independent, random all-to-all couplings $J_{ijkl}$.  Regarding the last property, Eqs.~\eqref{J_variance} and \eqref{Ndot_relation} imply that independence requires $\bar J \sim 1/N^3$ for a dot with local interactions.  Since $\bar J \ll \epsilon$ excited dot states indeed can be safely ignored.  However, increasing $N$ rapidly diminishes the strong-coupling temperature window $T \ll \bar J$---where much of the interesting physics emerges.  This challenge can be alleviated with long-range interactions, which lead to slower decay with $N$.  Alternatively, one can intentionally abandon independence to boost $\bar J$, though the fate of SYK physics in such cases remains to be systematically understood.

To maintain approximate $\mathcal{T}$ symmetry graphene-based dots appear ideal due to their strict two-dimensionality and extremely weak spin-orbit coupling. In this case the dominant source of $\mathcal{T}$ violation will likely originate from the Majorana wires. We can crudely assess the impact of such perturbations by adding local $\mathcal{T}$-breaking terms for the dot in the vicinity of the wires and projecting, e.g.,
\begin{eqnarray}
  \delta H &=& \chi \sum_{i = 1}^N(i c_{a_i}^\dagger c_{a_i + 1} + H.c.) \rightarrow \sum_{1\leq j < k \leq N} i M_{jk} \gamma_j' \gamma_k'
  \nonumber \\
  M_{jk} &=& \chi \sum_{i = 1}^N (\phi_{j,a_i}\phi_{k,a_i + 1} - \phi_{k,a_i} \phi_{j,a_i+1})
\end{eqnarray}
The $M_{jk}$ bilinear couplings are random with zero mean and variance $2\chi^2 N/N_{\rm dot}^2 \sim 1/N^5$, where we used Eq.~\eqref{Ndot_relation}.  The correction to the two-point correlation function $\langle \gamma_i(t) \gamma_i(0)\rangle$ is thus $\propto N(\chi^2/N^5)$, and should be compared to the contribution $\bar J^2 \sim U^2/N^6$ (for local interactions) from four-Majorana interactions.  This correction is small provided $\chi \lesssim U/N$; longer-range interactions relax the criterion further.  Tunneling into the dot provides an appealing benchmark of proximity to SYK physics: the conductance approaches a constant at zero bias if bilinears dominate but diverges as $V^{-1/2}$ for the large-$N$ SYK model \cite{Maldacena}.  

The setup we propose suggests several other tantalizing applications.  First, with relatively few wires ($N \leq 8$) one can experimentally explore the $\mathbb{Z} \rightarrow \mathbb{Z}_8$ reduction of the BDI classification by interactions \cite{FidkowskiKitaev1,FidkowskiKitaev2}, very similar to Refs.~\onlinecite{PikulinFranz1,PikulinFranz2}.  One can also investigate quantum quenches as a possible probe of SYK physics by disconnecting the dot and wires, effectively freezing the zero modes.  Finally, leveraging tools from topological quantum computation to measure quantities related to quantum information and chaos, e.g., out-of-time-order correlators or entanglement entropy, remains a particularly exciting prospect.

{\bf \emph{Acknowledgments.}}~We are indebted to X.~Chen, M.~Franz, Y.~Gu, A.~Kitaev, J.~Meyer, P.~Lee, S.~Nadj-Perge, J.~Iverson, and D.~Pikulin for illuminating discussions.  We gratefully acknowledge support from the National Science Foundation through grant DMR-1341822 (A.~C.~and J.\ A.); the Caltech Institute for Quantum Information and Matter, an NSF Physics Frontiers Center with support of the Gordon and Betty Moore Foundation through Grant GBMF1250; and the Walter Burke Institute for Theoretical Physics at Caltech.

\bibliography{SYK_references}

\begin{thebibliography}{58}%
\makeatletter
\providecommand \@ifxundefined [1]{%
 \@ifx{#1\undefined}
}%
\providecommand \@ifnum [1]{%
 \ifnum #1\expandafter \@firstoftwo
 \else \expandafter \@secondoftwo
 \fi
}%
\providecommand \@ifx [1]{%
 \ifx #1\expandafter \@firstoftwo
 \else \expandafter \@secondoftwo
 \fi
}%
\providecommand \natexlab [1]{#1}%
\providecommand \enquote  [1]{``#1''}%
\providecommand \bibnamefont  [1]{#1}%
\providecommand \bibfnamefont [1]{#1}%
\providecommand \citenamefont [1]{#1}%
\providecommand \href@noop [0]{\@secondoftwo}%
\providecommand \href [0]{\begingroup \@sanitize@url \@href}%
\providecommand \@href[1]{\@@startlink{#1}\@@href}%
\providecommand \@@href[1]{\endgroup#1\@@endlink}%
\providecommand \@sanitize@url [0]{\catcode `\\12\catcode `\$12\catcode
  `\&12\catcode `\#12\catcode `\^12\catcode `\_12\catcode `\%12\relax}%
\providecommand \@@startlink[1]{}%
\providecommand \@@endlink[0]{}%
\providecommand \url  [0]{\begingroup\@sanitize@url \@url }%
\providecommand \@url [1]{\endgroup\@href {#1}{\urlprefix }}%
\providecommand \urlprefix  [0]{URL }%
\providecommand \Eprint [0]{\href }%
\providecommand \doibase [0]{http://dx.doi.org/}%
\providecommand \selectlanguage [0]{\@gobble}%
\providecommand \bibinfo  [0]{\@secondoftwo}%
\providecommand \bibfield  [0]{\@secondoftwo}%
\providecommand \translation [1]{[#1]}%
\providecommand \BibitemOpen [0]{}%
\providecommand \bibitemStop [0]{}%
\providecommand \bibitemNoStop [0]{.\EOS\space}%
\providecommand \EOS [0]{\spacefactor3000\relax}%
\providecommand \BibitemShut  [1]{\csname bibitem#1\endcsname}%
\let\auto@bib@innerbib\@empty
%</preamble>
\bibitem [{\citenamefont {Read}\ and\ \citenamefont {Green}(2000)}]{ReadGreen}%
  \BibitemOpen
  \bibfield  {author} {\bibinfo {author} {\bibfnamefont {N.}~\bibnamefont
  {Read}}\ and\ \bibinfo {author} {\bibfnamefont {D.}~\bibnamefont {Green}},\
  }\href {\doibase 10.1103/PhysRevB.61.10267} {\bibfield  {journal} {\bibinfo
  {journal} {Phys. Rev. B}\ }\textbf {\bibinfo {volume} {61}},\ \bibinfo
  {pages} {10267} (\bibinfo {year} {2000})}\BibitemShut {NoStop}%
\bibitem [{\citenamefont {Kitaev}(2001)}]{1DwiresKitaev}%
  \BibitemOpen
  \bibfield  {author} {\bibinfo {author} {\bibfnamefont {A.~Y.}\ \bibnamefont
  {Kitaev}},\ }\href {\doibase 10.1070/1063-7869/44/10S/S29} {\bibfield
  {journal} {\bibinfo  {journal} {Sov. Phys.--Uspeki}\ }\textbf {\bibinfo
  {volume} {44}},\ \bibinfo {pages} {131} (\bibinfo {year} {2001})}\BibitemShut
  {NoStop}%
\bibitem [{\citenamefont {Kitaev}(2003)}]{kitaev}%
  \BibitemOpen
  \bibfield  {author} {\bibinfo {author} {\bibfnamefont {A.~Y.}\ \bibnamefont
  {Kitaev}},\ }\href {\doibase 10.1016/S0003-4916(02)00018-0} {\bibfield
  {journal} {\bibinfo  {journal} {Ann.\ Phys.}\ }\textbf {\bibinfo {volume}
  {303}},\ \bibinfo {pages} {2} (\bibinfo {year} {2003})}\BibitemShut {NoStop}%
\bibitem [{\citenamefont {Nayak}\ \emph {et~al.}(2008)\citenamefont {Nayak},
  \citenamefont {Simon}, \citenamefont {Stern}, \citenamefont {Freedman},\ and\
  \citenamefont {Das~Sarma}}]{TQCreview}%
  \BibitemOpen
  \bibfield  {author} {\bibinfo {author} {\bibfnamefont {C.}~\bibnamefont
  {Nayak}}, \bibinfo {author} {\bibfnamefont {S.~H.}\ \bibnamefont {Simon}},
  \bibinfo {author} {\bibfnamefont {A.}~\bibnamefont {Stern}}, \bibinfo
  {author} {\bibfnamefont {M.}~\bibnamefont {Freedman}}, \ and\ \bibinfo
  {author} {\bibfnamefont {S.}~\bibnamefont {Das~Sarma}},\ }\href {\doibase
  10.1103/RevModPhys.80.1083} {\bibfield  {journal} {\bibinfo  {journal} {Rev.
  Mod. Phys.}\ }\textbf {\bibinfo {volume} {80}},\ \bibinfo {pages} {1083}
  (\bibinfo {year} {2008})}\BibitemShut {NoStop}%
\bibitem [{\citenamefont {Mourik}\ \emph {et~al.}(2012)\citenamefont {Mourik},
  \citenamefont {Zuo}, \citenamefont {Frolov}, \citenamefont {Plissard},
  \citenamefont {Bakkers},\ and\ \citenamefont {Kouwenhoven}}]{mourik12}%
  \BibitemOpen
  \bibfield  {author} {\bibinfo {author} {\bibfnamefont {V.}~\bibnamefont
  {Mourik}}, \bibinfo {author} {\bibfnamefont {K.}~\bibnamefont {Zuo}},
  \bibinfo {author} {\bibfnamefont {S.~M.}\ \bibnamefont {Frolov}}, \bibinfo
  {author} {\bibfnamefont {S.~R.}\ \bibnamefont {Plissard}}, \bibinfo {author}
  {\bibfnamefont {E.~P. A.~M.}\ \bibnamefont {Bakkers}}, \ and\ \bibinfo
  {author} {\bibfnamefont {L.~P.}\ \bibnamefont {Kouwenhoven}},\ }\href
  {\doibase 10.1126/science.1222360} {\bibfield  {journal} {\bibinfo  {journal}
  {Science}\ }\textbf {\bibinfo {volume} {336}},\ \bibinfo {pages} {1003}
  (\bibinfo {year} {2012})}\BibitemShut {NoStop}%
\bibitem [{\citenamefont {Das}\ \emph {et~al.}(2012)\citenamefont {Das},
  \citenamefont {Ronen}, \citenamefont {Most}, \citenamefont {Oreg},
  \citenamefont {Heiblum},\ and\ \citenamefont {Shtrikman}}]{das12}%
  \BibitemOpen
  \bibfield  {author} {\bibinfo {author} {\bibfnamefont {A.}~\bibnamefont
  {Das}}, \bibinfo {author} {\bibfnamefont {Y.}~\bibnamefont {Ronen}}, \bibinfo
  {author} {\bibfnamefont {Y.}~\bibnamefont {Most}}, \bibinfo {author}
  {\bibfnamefont {Y.}~\bibnamefont {Oreg}}, \bibinfo {author} {\bibfnamefont
  {M.}~\bibnamefont {Heiblum}}, \ and\ \bibinfo {author} {\bibfnamefont
  {H.}~\bibnamefont {Shtrikman}},\ }\href {\doibase 10.1038/nphys2479}
  {\bibfield  {journal} {\bibinfo  {journal} {Nat. Phys.}\ }\textbf {\bibinfo
  {volume} {8}},\ \bibinfo {pages} {887} (\bibinfo {year} {2012})}\BibitemShut
  {NoStop}%
\bibitem [{\citenamefont {Finck}\ \emph {et~al.}(2013)\citenamefont {Finck},
  \citenamefont {Van~Harlingen}, \citenamefont {Mohseni}, \citenamefont
  {Jung},\ and\ \citenamefont {Li}}]{finck12}%
  \BibitemOpen
  \bibfield  {author} {\bibinfo {author} {\bibfnamefont {A.~D.~K.}\
  \bibnamefont {Finck}}, \bibinfo {author} {\bibfnamefont {D.~J.}\ \bibnamefont
  {Van~Harlingen}}, \bibinfo {author} {\bibfnamefont {P.~K.}\ \bibnamefont
  {Mohseni}}, \bibinfo {author} {\bibfnamefont {K.}~\bibnamefont {Jung}}, \
  and\ \bibinfo {author} {\bibfnamefont {X.}~\bibnamefont {Li}},\ }\href
  {\doibase 10.1103/PhysRevLett.110.126406} {\bibfield  {journal} {\bibinfo
  {journal} {Phys. Rev. Lett.}\ }\textbf {\bibinfo {volume} {110}},\ \bibinfo
  {pages} {126406} (\bibinfo {year} {2013})}\BibitemShut {NoStop}%
\bibitem [{\citenamefont {Churchill}\ \emph {et~al.}(2013)\citenamefont
  {Churchill}, \citenamefont {Fatemi}, \citenamefont {Grove-Rasmussen},
  \citenamefont {Deng}, \citenamefont {Caroff}, \citenamefont {Xu},\ and\
  \citenamefont {Marcus}}]{Churchill}%
  \BibitemOpen
  \bibfield  {author} {\bibinfo {author} {\bibfnamefont {H.~O.~H.}\
  \bibnamefont {Churchill}}, \bibinfo {author} {\bibfnamefont {V.}~\bibnamefont
  {Fatemi}}, \bibinfo {author} {\bibfnamefont {K.}~\bibnamefont
  {Grove-Rasmussen}}, \bibinfo {author} {\bibfnamefont {M.~T.}\ \bibnamefont
  {Deng}}, \bibinfo {author} {\bibfnamefont {P.}~\bibnamefont {Caroff}},
  \bibinfo {author} {\bibfnamefont {H.~Q.}\ \bibnamefont {Xu}}, \ and\ \bibinfo
  {author} {\bibfnamefont {C.~M.}\ \bibnamefont {Marcus}},\ }\href {\doibase
  10.1103/PhysRevB.87.241401} {\bibfield  {journal} {\bibinfo  {journal} {Phys.
  Rev. B}\ }\textbf {\bibinfo {volume} {87}},\ \bibinfo {pages} {241401}
  (\bibinfo {year} {2013})}\BibitemShut {NoStop}%
\bibitem [{\citenamefont {Lee}\ \emph {et~al.}(2014)\citenamefont {Lee},
  \citenamefont {Jiang}, \citenamefont {Houzet}, \citenamefont {Aguado},
  \citenamefont {Lieber},\ and\ \citenamefont {{De
  Franceschi}}}]{DeFranceschi}%
  \BibitemOpen
  \bibfield  {author} {\bibinfo {author} {\bibfnamefont {E.~J.~H.}\
  \bibnamefont {Lee}}, \bibinfo {author} {\bibfnamefont {X.}~\bibnamefont
  {Jiang}}, \bibinfo {author} {\bibfnamefont {M.}~\bibnamefont {Houzet}},
  \bibinfo {author} {\bibfnamefont {R.}~\bibnamefont {Aguado}}, \bibinfo
  {author} {\bibfnamefont {C.~M.}\ \bibnamefont {Lieber}}, \ and\ \bibinfo
  {author} {\bibfnamefont {S.}~\bibnamefont {{De Franceschi}}},\ }\href
  {\doibase 10.1038/nnano.2013.267} {\bibfield  {journal} {\bibinfo  {journal}
  {Nature Nanotech.}\ }\textbf {\bibinfo {volume} {9}},\ \bibinfo {pages} {79}
  (\bibinfo {year} {2014})}\BibitemShut {NoStop}%
\bibitem [{\citenamefont {Nadj-Perge}\ \emph {et~al.}(2014)\citenamefont
  {Nadj-Perge}, \citenamefont {Drozdov}, \citenamefont {Li}, \citenamefont
  {Chen}, \citenamefont {Jeon}, \citenamefont {Seo}, \citenamefont {MacDonald},
  \citenamefont {Bernevig},\ and\ \citenamefont {Yazdani}}]{Nadj-Perge}%
  \BibitemOpen
  \bibfield  {author} {\bibinfo {author} {\bibfnamefont {S.}~\bibnamefont
  {Nadj-Perge}}, \bibinfo {author} {\bibfnamefont {I.~K.}\ \bibnamefont
  {Drozdov}}, \bibinfo {author} {\bibfnamefont {J.}~\bibnamefont {Li}},
  \bibinfo {author} {\bibfnamefont {H.}~\bibnamefont {Chen}}, \bibinfo {author}
  {\bibfnamefont {S.}~\bibnamefont {Jeon}}, \bibinfo {author} {\bibfnamefont
  {J.}~\bibnamefont {Seo}}, \bibinfo {author} {\bibfnamefont {A.~H.}\
  \bibnamefont {MacDonald}}, \bibinfo {author} {\bibfnamefont {B.~A.}\
  \bibnamefont {Bernevig}}, \ and\ \bibinfo {author} {\bibfnamefont
  {A.}~\bibnamefont {Yazdani}},\ }\href {\doibase 10.1126/science.1259327}
  {\bibfield  {journal} {\bibinfo  {journal} {Science}\ }\textbf {\bibinfo
  {volume} {346}},\ \bibinfo {pages} {602} (\bibinfo {year}
  {2014})}\BibitemShut {NoStop}%
\bibitem [{\citenamefont {Kurter}\ \emph {et~al.}(2015)\citenamefont {Kurter},
  \citenamefont {Finck}, \citenamefont {Hor},\ and\ \citenamefont {{Van
  Harlingen}}}]{Kurter}%
  \BibitemOpen
  \bibfield  {author} {\bibinfo {author} {\bibfnamefont {C.}~\bibnamefont
  {Kurter}}, \bibinfo {author} {\bibfnamefont {A.~D.~K.}\ \bibnamefont
  {Finck}}, \bibinfo {author} {\bibfnamefont {Y.~S.}\ \bibnamefont {Hor}}, \
  and\ \bibinfo {author} {\bibfnamefont {D.~J.}\ \bibnamefont {{Van
  Harlingen}}},\ }\href {\doibase 10.1038/ncomms8130} {\bibfield  {journal}
  {\bibinfo  {journal} {Nature Communications}\ }\textbf {\bibinfo {volume}
  {6}},\ \bibinfo {pages} {7130} (\bibinfo {year} {2015})}\BibitemShut
  {NoStop}%
\bibitem [{\citenamefont {Xu}\ \emph {et~al.}(2015)\citenamefont {Xu},
  \citenamefont {Wang}, \citenamefont {Liu}, \citenamefont {Ge}, \citenamefont
  {Yang}, \citenamefont {Liu}, \citenamefont {Xu}, \citenamefont {Guan},
  \citenamefont {Gao}, \citenamefont {Qian}, \citenamefont {Liu}, \citenamefont
  {Wang}, \citenamefont {Zhang}, \citenamefont {Xue},\ and\ \citenamefont
  {Jia}}]{Xu}%
  \BibitemOpen
  \bibfield  {author} {\bibinfo {author} {\bibfnamefont {J.-P.}\ \bibnamefont
  {Xu}}, \bibinfo {author} {\bibfnamefont {M.-X.}\ \bibnamefont {Wang}},
  \bibinfo {author} {\bibfnamefont {Z.~L.}\ \bibnamefont {Liu}}, \bibinfo
  {author} {\bibfnamefont {J.-F.}\ \bibnamefont {Ge}}, \bibinfo {author}
  {\bibfnamefont {X.}~\bibnamefont {Yang}}, \bibinfo {author} {\bibfnamefont
  {C.}~\bibnamefont {Liu}}, \bibinfo {author} {\bibfnamefont {Z.~A.}\
  \bibnamefont {Xu}}, \bibinfo {author} {\bibfnamefont {D.}~\bibnamefont
  {Guan}}, \bibinfo {author} {\bibfnamefont {C.~L.}\ \bibnamefont {Gao}},
  \bibinfo {author} {\bibfnamefont {D.}~\bibnamefont {Qian}}, \bibinfo {author}
  {\bibfnamefont {Y.}~\bibnamefont {Liu}}, \bibinfo {author} {\bibfnamefont
  {Q.-H.}\ \bibnamefont {Wang}}, \bibinfo {author} {\bibfnamefont {F.-C.}\
  \bibnamefont {Zhang}}, \bibinfo {author} {\bibfnamefont {Q.-K.}\ \bibnamefont
  {Xue}}, \ and\ \bibinfo {author} {\bibfnamefont {J.-F.}\ \bibnamefont
  {Jia}},\ }\href {\doibase 10.1103/PhysRevLett.114.017001} {\bibfield
  {journal} {\bibinfo  {journal} {Phys. Rev. Lett.}\ }\textbf {\bibinfo
  {volume} {114}},\ \bibinfo {pages} {017001} (\bibinfo {year}
  {2015})}\BibitemShut {NoStop}%
\bibitem [{\citenamefont {{Albrecht}}\ \emph {et~al.}(2016)\citenamefont
  {{Albrecht}}, \citenamefont {{Higginbotham}}, \citenamefont {{Madsen}},
  \citenamefont {{Kuemmeth}}, \citenamefont {{Jespersen}}, \citenamefont
  {{Nyg{\aa}rd}}, \citenamefont {{Krogstrup}},\ and\ \citenamefont
  {{Marcus}}}]{AlbrechtExponential}%
  \BibitemOpen
  \bibfield  {author} {\bibinfo {author} {\bibfnamefont {S.~M.}\ \bibnamefont
  {{Albrecht}}}, \bibinfo {author} {\bibfnamefont {A.~P.}\ \bibnamefont
  {{Higginbotham}}}, \bibinfo {author} {\bibfnamefont {M.}~\bibnamefont
  {{Madsen}}}, \bibinfo {author} {\bibfnamefont {F.}~\bibnamefont
  {{Kuemmeth}}}, \bibinfo {author} {\bibfnamefont {T.~S.}\ \bibnamefont
  {{Jespersen}}}, \bibinfo {author} {\bibfnamefont {J.}~\bibnamefont
  {{Nyg{\aa}rd}}}, \bibinfo {author} {\bibfnamefont {P.}~\bibnamefont
  {{Krogstrup}}}, \ and\ \bibinfo {author} {\bibfnamefont {C.~M.}\ \bibnamefont
  {{Marcus}}},\ }\href {\doibase 10.1038/nature17162} {\bibfield  {journal}
  {\bibinfo  {journal} {Nature}\ }\textbf {\bibinfo {volume} {531}},\ \bibinfo
  {pages} {206} (\bibinfo {year} {2016})}\BibitemShut {NoStop}%
\bibitem [{\citenamefont {Zhang}\ \emph {et~al.}(2016)\citenamefont {Zhang},
  \citenamefont {Gul}, \citenamefont {Conesa-Boj}, \citenamefont {Zuo},
  \citenamefont {Mourik}, \citenamefont {{de Vries}}, \citenamefont {{van
  Veen}}, \citenamefont {{van Woerkom}}, \citenamefont {Nowak}, \citenamefont
  {Wimmer}, \citenamefont {Car}, \citenamefont {Plissard}, \citenamefont
  {Bakkers}, \citenamefont {Quintero-Perez}, \citenamefont {Goswami},
  \citenamefont {Watanabe}, \citenamefont {Taniguchi},\ and\ \citenamefont
  {Kouwenhoven}}]{Gul2}%
  \BibitemOpen
  \bibfield  {author} {\bibinfo {author} {\bibfnamefont {H.}~\bibnamefont
  {Zhang}}, \bibinfo {author} {\bibfnamefont {O.}~\bibnamefont {Gul}}, \bibinfo
  {author} {\bibfnamefont {S.}~\bibnamefont {Conesa-Boj}}, \bibinfo {author}
  {\bibfnamefont {K.}~\bibnamefont {Zuo}}, \bibinfo {author} {\bibfnamefont
  {V.}~\bibnamefont {Mourik}}, \bibinfo {author} {\bibfnamefont {F.~K.}\
  \bibnamefont {{de Vries}}}, \bibinfo {author} {\bibfnamefont
  {J.}~\bibnamefont {{van Veen}}}, \bibinfo {author} {\bibfnamefont {D.~J.}\
  \bibnamefont {{van Woerkom}}}, \bibinfo {author} {\bibfnamefont {M.~P.}\
  \bibnamefont {Nowak}}, \bibinfo {author} {\bibfnamefont {M.}~\bibnamefont
  {Wimmer}}, \bibinfo {author} {\bibfnamefont {D.}~\bibnamefont {Car}},
  \bibinfo {author} {\bibfnamefont {S.}~\bibnamefont {Plissard}}, \bibinfo
  {author} {\bibfnamefont {E.~P. A.~M.}\ \bibnamefont {Bakkers}}, \bibinfo
  {author} {\bibfnamefont {M.}~\bibnamefont {Quintero-Perez}}, \bibinfo
  {author} {\bibfnamefont {S.}~\bibnamefont {Goswami}}, \bibinfo {author}
  {\bibfnamefont {K.}~\bibnamefont {Watanabe}}, \bibinfo {author}
  {\bibfnamefont {T.}~\bibnamefont {Taniguchi}}, \ and\ \bibinfo {author}
  {\bibfnamefont {L.~P.}\ \bibnamefont {Kouwenhoven}},\ }\href@noop {}
  {\bibfield  {journal} {\bibinfo  {journal} {arXiv:1603.04069}\ } (\bibinfo
  {year} {2016})}\BibitemShut {NoStop}%
\bibitem [{\citenamefont {Deng}\ \emph {et~al.}(2016)\citenamefont {Deng},
  \citenamefont {Vaitiekenas}, \citenamefont {Hansen}, \citenamefont {Danon},
  \citenamefont {Leijnse}, \citenamefont {Flensberg}, \citenamefont {Nyg{\r
  a}rd}, \citenamefont {Krogstrup},\ and\ \citenamefont
  {Marcus}}]{DengPhaseTransition}%
  \BibitemOpen
  \bibfield  {author} {\bibinfo {author} {\bibfnamefont {M.~T.}\ \bibnamefont
  {Deng}}, \bibinfo {author} {\bibfnamefont {S.}~\bibnamefont {Vaitiekenas}},
  \bibinfo {author} {\bibfnamefont {E.~B.}\ \bibnamefont {Hansen}}, \bibinfo
  {author} {\bibfnamefont {J.}~\bibnamefont {Danon}}, \bibinfo {author}
  {\bibfnamefont {M.}~\bibnamefont {Leijnse}}, \bibinfo {author} {\bibfnamefont
  {K.}~\bibnamefont {Flensberg}}, \bibinfo {author} {\bibfnamefont
  {J.}~\bibnamefont {Nyg{\r a}rd}}, \bibinfo {author} {\bibfnamefont
  {P.}~\bibnamefont {Krogstrup}}, \ and\ \bibinfo {author} {\bibfnamefont
  {C.~M.}\ \bibnamefont {Marcus}},\ }\href {\doibase 10.1126/science.aaf3961}
  {\bibfield  {journal} {\bibinfo  {journal} {Science}\ }\textbf {\bibinfo
  {volume} {354}},\ \bibinfo {pages} {1557} (\bibinfo {year}
  {2016})}\BibitemShut {NoStop}%
\bibitem [{\citenamefont {Feldman}\ \emph {et~al.}(2017)\citenamefont
  {Feldman}, \citenamefont {Randeria}, \citenamefont {Li}, \citenamefont
  {Jeon}, \citenamefont {Xie}, \citenamefont {Wang}, \citenamefont {Drozdov},
  \citenamefont {Bernevig},\ and\ \citenamefont {Yazdani}}]{Chains2}%
  \BibitemOpen
  \bibfield  {author} {\bibinfo {author} {\bibfnamefont {B.~E.}\ \bibnamefont
  {Feldman}}, \bibinfo {author} {\bibfnamefont {M.~T.}\ \bibnamefont
  {Randeria}}, \bibinfo {author} {\bibfnamefont {J.}~\bibnamefont {Li}},
  \bibinfo {author} {\bibfnamefont {S.}~\bibnamefont {Jeon}}, \bibinfo {author}
  {\bibfnamefont {Y.}~\bibnamefont {Xie}}, \bibinfo {author} {\bibfnamefont
  {Z.}~\bibnamefont {Wang}}, \bibinfo {author} {\bibfnamefont {I.~K.}\
  \bibnamefont {Drozdov}}, \bibinfo {author} {\bibfnamefont {B.~A.}\
  \bibnamefont {Bernevig}}, \ and\ \bibinfo {author} {\bibfnamefont
  {A.}~\bibnamefont {Yazdani}},\ }\href {\doibase 10.1038/nphys3947} {\bibfield
   {journal} {\bibinfo  {journal} {Nature Physics}\ }\textbf {\bibinfo {volume}
  {13}},\ \bibinfo {pages} {286} (\bibinfo {year} {2017})}\BibitemShut
  {NoStop}%
\bibitem [{\citenamefont {Sachdev}\ and\ \citenamefont {Ye}(1993)}]{Sachdev}%
  \BibitemOpen
  \bibfield  {author} {\bibinfo {author} {\bibfnamefont {S.}~\bibnamefont
  {Sachdev}}\ and\ \bibinfo {author} {\bibfnamefont {J.}~\bibnamefont {Ye}},\
  }\href {\doibase 10.1103/PhysRevLett.70.3339} {\bibfield  {journal} {\bibinfo
   {journal} {Phys. Rev. Lett.}\ }\textbf {\bibinfo {volume} {70}},\ \bibinfo
  {pages} {3339} (\bibinfo {year} {1993})}\BibitemShut {NoStop}%
\bibitem [{\citenamefont {Kitaev}(2015)}]{KitaevTalks}%
  \BibitemOpen
  \bibfield  {author} {\bibinfo {author} {\bibfnamefont {A.}~\bibnamefont
  {Kitaev}},\ }\href@noop {} {\bibfield  {journal} {\bibinfo  {journal}
  {http://online.kitp.ucsb.edu/online/entangled15/ kitaev/,
  http://online.kitp.ucsb.edu/online/entangled15/ kitaev2/}\ } (\bibinfo {year}
  {2015})}\BibitemShut {NoStop}%
\bibitem [{\citenamefont {Maldacena}\ and\ \citenamefont
  {Stanford}(2016)}]{Maldacena}%
  \BibitemOpen
  \bibfield  {author} {\bibinfo {author} {\bibfnamefont {J.}~\bibnamefont
  {Maldacena}}\ and\ \bibinfo {author} {\bibfnamefont {D.}~\bibnamefont
  {Stanford}},\ }\href {\doibase 10.1103/PhysRevD.94.106002} {\bibfield
  {journal} {\bibinfo  {journal} {Phys. Rev. D}\ }\textbf {\bibinfo {volume}
  {94}},\ \bibinfo {pages} {106002} (\bibinfo {year} {2016})}\BibitemShut
  {NoStop}%
\bibitem [{\citenamefont {Polchinski}\ and\ \citenamefont
  {Rosenhaus}(2016)}]{Polchinski}%
  \BibitemOpen
  \bibfield  {author} {\bibinfo {author} {\bibfnamefont {J.}~\bibnamefont
  {Polchinski}}\ and\ \bibinfo {author} {\bibfnamefont {V.}~\bibnamefont
  {Rosenhaus}},\ }\href@noop {} {\bibfield  {journal} {\bibinfo  {journal}
  {arXiv:1601.06768}\ } (\bibinfo {year} {2016})}\BibitemShut {NoStop}%
\bibitem [{\citenamefont {You}\ \emph {et~al.}(2016)\citenamefont {You},
  \citenamefont {Ludwig},\ and\ \citenamefont {Xu}}]{YouSYK}%
  \BibitemOpen
  \bibfield  {author} {\bibinfo {author} {\bibfnamefont {Y.-Z.}\ \bibnamefont
  {You}}, \bibinfo {author} {\bibfnamefont {A.~W.~W.}\ \bibnamefont {Ludwig}},
  \ and\ \bibinfo {author} {\bibfnamefont {C.}~\bibnamefont {Xu}},\ }\href@noop
  {} {\bibfield  {journal} {\bibinfo  {journal} {arXiv:1602.06964}\ } (\bibinfo
  {year} {2016})}\BibitemShut {NoStop}%
\bibitem [{\citenamefont {Fu}\ and\ \citenamefont {Sachdev}(2016)}]{FuSYK}%
  \BibitemOpen
  \bibfield  {author} {\bibinfo {author} {\bibfnamefont {W.}~\bibnamefont
  {Fu}}\ and\ \bibinfo {author} {\bibfnamefont {S.}~\bibnamefont {Sachdev}},\
  }\href {\doibase 10.1103/PhysRevB.94.035135} {\bibfield  {journal} {\bibinfo
  {journal} {Phys. Rev. B}\ }\textbf {\bibinfo {volume} {94}},\ \bibinfo
  {pages} {035135} (\bibinfo {year} {2016})}\BibitemShut {NoStop}%
\bibitem [{\citenamefont {Danshita}\ \emph {et~al.}(2016)\citenamefont
  {Danshita}, \citenamefont {Hanada},\ and\ \citenamefont {Tezuka}}]{Danshita}%
  \BibitemOpen
  \bibfield  {author} {\bibinfo {author} {\bibfnamefont {I.}~\bibnamefont
  {Danshita}}, \bibinfo {author} {\bibfnamefont {M.}~\bibnamefont {Hanada}}, \
  and\ \bibinfo {author} {\bibfnamefont {M.}~\bibnamefont {Tezuka}},\
  }\href@noop {} {\bibfield  {journal} {\bibinfo  {journal} {arXiv:1606.02454}\
  } (\bibinfo {year} {2016})}\BibitemShut {NoStop}%
\bibitem [{\citenamefont {Jevicki}\ and\ \citenamefont
  {Suzuki}(2016)}]{Jevicki}%
  \BibitemOpen
  \bibfield  {author} {\bibinfo {author} {\bibfnamefont {A.}~\bibnamefont
  {Jevicki}}\ and\ \bibinfo {author} {\bibfnamefont {K.}~\bibnamefont
  {Suzuki}},\ }\href@noop {} {\bibfield  {journal} {\bibinfo  {journal}
  {arXiv:1608.07567}\ } (\bibinfo {year} {2016})}\BibitemShut {NoStop}%
\bibitem [{\citenamefont {Gross}\ and\ \citenamefont
  {Rosenhaus}(2016)}]{Gross}%
  \BibitemOpen
  \bibfield  {author} {\bibinfo {author} {\bibfnamefont {D.~J.}\ \bibnamefont
  {Gross}}\ and\ \bibinfo {author} {\bibfnamefont {V.}~\bibnamefont
  {Rosenhaus}},\ }\href@noop {} {\bibfield  {journal} {\bibinfo  {journal}
  {arXiv:1610.01569}\ } (\bibinfo {year} {2016})}\BibitemShut {NoStop}%
\bibitem [{\citenamefont {Banerjee}\ and\ \citenamefont
  {Altman}(2016)}]{Banerjee}%
  \BibitemOpen
  \bibfield  {author} {\bibinfo {author} {\bibfnamefont {S.}~\bibnamefont
  {Banerjee}}\ and\ \bibinfo {author} {\bibfnamefont {E.}~\bibnamefont
  {Altman}},\ }\href@noop {} {\bibfield  {journal} {\bibinfo  {journal}
  {arXiv:1610.04619}\ } (\bibinfo {year} {2016})}\BibitemShut {NoStop}%
\bibitem [{\citenamefont {Witten}(2016)}]{WittenSYK}%
  \BibitemOpen
  \bibfield  {author} {\bibinfo {author} {\bibfnamefont {E.}~\bibnamefont
  {Witten}},\ }\href@noop {} {\  (\bibinfo {year} {2016})},\ \Eprint
  {http://arxiv.org/abs/arXiv:1610.09758} {arXiv:1610.09758} \BibitemShut
  {NoStop}%
\bibitem [{\citenamefont {Gurau}(2017)}]{GurauSYK}%
  \BibitemOpen
  \bibfield  {author} {\bibinfo {author} {\bibfnamefont {R.}~\bibnamefont
  {Gurau}},\ }\href {\doibase
  http://dx.doi.org/10.1016/j.nuclphysb.2017.01.015} {\bibfield  {journal}
  {\bibinfo  {journal} {Nuclear Physics B}\ }\textbf {\bibinfo {volume}
  {916}},\ \bibinfo {pages} {386 } (\bibinfo {year} {2017})}\BibitemShut
  {NoStop}%
\bibitem [{\citenamefont {Cotler}\ \emph {et~al.}(2016)\citenamefont {Cotler},
  \citenamefont {Gur-Ari}, \citenamefont {Hanada}, \citenamefont {Polchinski},
  \citenamefont {Saad}, \citenamefont {Shenker}, \citenamefont {Stanford},
  \citenamefont {Streicher},\ and\ \citenamefont {Tezuka}}]{ColterSYK}%
  \BibitemOpen
  \bibfield  {author} {\bibinfo {author} {\bibfnamefont {J.~S.}\ \bibnamefont
  {Cotler}}, \bibinfo {author} {\bibfnamefont {G.}~\bibnamefont {Gur-Ari}},
  \bibinfo {author} {\bibfnamefont {M.}~\bibnamefont {Hanada}}, \bibinfo
  {author} {\bibfnamefont {J.}~\bibnamefont {Polchinski}}, \bibinfo {author}
  {\bibfnamefont {P.}~\bibnamefont {Saad}}, \bibinfo {author} {\bibfnamefont
  {S.~H.}\ \bibnamefont {Shenker}}, \bibinfo {author} {\bibfnamefont
  {D.}~\bibnamefont {Stanford}}, \bibinfo {author} {\bibfnamefont
  {A.}~\bibnamefont {Streicher}}, \ and\ \bibinfo {author} {\bibfnamefont
  {M.}~\bibnamefont {Tezuka}},\ }\href@noop {} {\  (\bibinfo {year} {2016})},\
  \Eprint {http://arxiv.org/abs/arXiv:1611.04650} {arXiv:1611.04650}
  \BibitemShut {NoStop}%
\bibitem [{\citenamefont {Davison}\ \emph {et~al.}(2016)\citenamefont
  {Davison}, \citenamefont {Fu}, \citenamefont {Georges}, \citenamefont {Gu},
  \citenamefont {Jensen},\ and\ \citenamefont {Sachdev}}]{DavisonSYK}%
  \BibitemOpen
  \bibfield  {author} {\bibinfo {author} {\bibfnamefont {R.~A.}\ \bibnamefont
  {Davison}}, \bibinfo {author} {\bibfnamefont {W.}~\bibnamefont {Fu}},
  \bibinfo {author} {\bibfnamefont {A.}~\bibnamefont {Georges}}, \bibinfo
  {author} {\bibfnamefont {Y.}~\bibnamefont {Gu}}, \bibinfo {author}
  {\bibfnamefont {K.}~\bibnamefont {Jensen}}, \ and\ \bibinfo {author}
  {\bibfnamefont {S.}~\bibnamefont {Sachdev}},\ }\href@noop {} {\  (\bibinfo
  {year} {2016})},\ \Eprint {http://arxiv.org/abs/arXiv:1612.00849}
  {arXiv:1612.00849} \BibitemShut {NoStop}%
\bibitem [{\citenamefont {Jensen}(2016)}]{JensenSYK}%
  \BibitemOpen
  \bibfield  {author} {\bibinfo {author} {\bibfnamefont {K.}~\bibnamefont
  {Jensen}},\ }\href {\doibase 10.1103/PhysRevLett.117.111601} {\bibfield
  {journal} {\bibinfo  {journal} {Phys. Rev. Lett.}\ }\textbf {\bibinfo
  {volume} {117}},\ \bibinfo {pages} {111601} (\bibinfo {year}
  {2016})}\BibitemShut {NoStop}%
\bibitem [{\citenamefont {Bi}\ \emph {et~al.}(2017)\citenamefont {Bi},
  \citenamefont {Jian}, \citenamefont {You}, \citenamefont {Pawlak},\ and\
  \citenamefont {Xu}}]{BiSYK}%
  \BibitemOpen
  \bibfield  {author} {\bibinfo {author} {\bibfnamefont {Z.}~\bibnamefont
  {Bi}}, \bibinfo {author} {\bibfnamefont {C.-M.}\ \bibnamefont {Jian}},
  \bibinfo {author} {\bibfnamefont {Y.-Z.}\ \bibnamefont {You}}, \bibinfo
  {author} {\bibfnamefont {K.~A.}\ \bibnamefont {Pawlak}}, \ and\ \bibinfo
  {author} {\bibfnamefont {C.}~\bibnamefont {Xu}},\ }\href@noop {} {\
  (\bibinfo {year} {2017})},\ \Eprint {http://arxiv.org/abs/arXiv:1701.07081}
  {arXiv:1701.07081} \BibitemShut {NoStop}%
\bibitem [{\citenamefont {Gu}\ \emph {et~al.}(2016)\citenamefont {Gu},
  \citenamefont {Qi},\ and\ \citenamefont {Stanford}}]{Gu}%
  \BibitemOpen
  \bibfield  {author} {\bibinfo {author} {\bibfnamefont {Y.}~\bibnamefont
  {Gu}}, \bibinfo {author} {\bibfnamefont {X.-L.}\ \bibnamefont {Qi}}, \ and\
  \bibinfo {author} {\bibfnamefont {D.}~\bibnamefont {Stanford}},\ }\href@noop
  {} {\bibfield  {journal} {\bibinfo  {journal} {arXiv:1609.07832}\ } (\bibinfo
  {year} {2016})}\BibitemShut {NoStop}%
\bibitem [{\citenamefont {Pikulin}\ and\ \citenamefont
  {Franz}(2017)}]{MarcelSYK}%
  \BibitemOpen
  \bibfield  {author} {\bibinfo {author} {\bibfnamefont {D.~I.}\ \bibnamefont
  {Pikulin}}\ and\ \bibinfo {author} {\bibfnamefont {M.}~\bibnamefont
  {Franz}},\ }\href@noop {} {\bibfield  {journal} {\bibinfo  {journal}
  {arXiv:1702.04426}\ } (\bibinfo {year} {2017})}\BibitemShut {NoStop}%
\bibitem [{\citenamefont {Gross}\ and\ \citenamefont
  {Rosenhaus}(2017)}]{Gross2}%
  \BibitemOpen
  \bibfield  {author} {\bibinfo {author} {\bibfnamefont {D.~J.}\ \bibnamefont
  {Gross}}\ and\ \bibinfo {author} {\bibfnamefont {V.}~\bibnamefont
  {Rosenhaus}},\ }\href@noop {} {\  (\bibinfo {year} {2017})},\ \Eprint
  {http://arxiv.org/abs/arXiv:1702.08016} {arXiv:1702.08016} \BibitemShut
  {NoStop}%
\bibitem [{\citenamefont {Lutchyn}\ \emph {et~al.}(2010)\citenamefont
  {Lutchyn}, \citenamefont {Sau},\ and\ \citenamefont
  {Das~Sarma}}]{1DwiresLutchyn}%
  \BibitemOpen
  \bibfield  {author} {\bibinfo {author} {\bibfnamefont {R.~M.}\ \bibnamefont
  {Lutchyn}}, \bibinfo {author} {\bibfnamefont {J.~D.}\ \bibnamefont {Sau}}, \
  and\ \bibinfo {author} {\bibfnamefont {S.}~\bibnamefont {Das~Sarma}},\ }\href
  {\doibase 10.1103/PhysRevLett.105.077001} {\bibfield  {journal} {\bibinfo
  {journal} {Phys.\ Rev.\ Lett.}\ }\textbf {\bibinfo {volume} {105}},\ \bibinfo
  {pages} {077001} (\bibinfo {year} {2010})}\BibitemShut {NoStop}%
\bibitem [{\citenamefont {Oreg}\ \emph {et~al.}(2010)\citenamefont {Oreg},
  \citenamefont {Refael},\ and\ \citenamefont {{von Oppen}}}]{1DwiresOreg}%
  \BibitemOpen
  \bibfield  {author} {\bibinfo {author} {\bibfnamefont {Y.}~\bibnamefont
  {Oreg}}, \bibinfo {author} {\bibfnamefont {G.}~\bibnamefont {Refael}}, \ and\
  \bibinfo {author} {\bibfnamefont {F.}~\bibnamefont {{von Oppen}}},\ }\href
  {\doibase 10.1103/PhysRevLett.105.177002} {\bibfield  {journal} {\bibinfo
  {journal} {Phys.\ Rev.\ Lett.}\ }\textbf {\bibinfo {volume} {105}},\ \bibinfo
  {pages} {177002} (\bibinfo {year} {2010})}\BibitemShut {NoStop}%
\bibitem [{\citenamefont {Tewari}\ and\ \citenamefont
  {Sau}(2012)}]{TewariInvariant}%
  \BibitemOpen
  \bibfield  {author} {\bibinfo {author} {\bibfnamefont {S.}~\bibnamefont
  {Tewari}}\ and\ \bibinfo {author} {\bibfnamefont {J.~D.}\ \bibnamefont
  {Sau}},\ }\href {\doibase 10.1103/PhysRevLett.109.150408} {\bibfield
  {journal} {\bibinfo  {journal} {Phys. Rev. Lett.}\ }\textbf {\bibinfo
  {volume} {109}},\ \bibinfo {pages} {150408} (\bibinfo {year}
  {2012})}\BibitemShut {NoStop}%
\bibitem [{\citenamefont {Kitaev}(2009)}]{KitaevClassification}%
  \BibitemOpen
  \bibfield  {author} {\bibinfo {author} {\bibfnamefont {A.}~\bibnamefont
  {Kitaev}},\ }\href {\doibase 10.1063/1.3149495} {\bibfield  {journal}
  {\bibinfo  {journal} {AIP Conference Proceedings}\ }\textbf {\bibinfo
  {volume} {1134}},\ \bibinfo {pages} {22} (\bibinfo {year}
  {2009})}\BibitemShut {NoStop}%
\bibitem [{\citenamefont {Ryu}\ \emph {et~al.}(2010)\citenamefont {Ryu},
  \citenamefont {Schnyder}, \citenamefont {Furusaki},\ and\ \citenamefont
  {Ludwig}}]{RyuClassification}%
  \BibitemOpen
  \bibfield  {author} {\bibinfo {author} {\bibfnamefont {S.}~\bibnamefont
  {Ryu}}, \bibinfo {author} {\bibfnamefont {A.~P.}\ \bibnamefont {Schnyder}},
  \bibinfo {author} {\bibfnamefont {A.}~\bibnamefont {Furusaki}}, \ and\
  \bibinfo {author} {\bibfnamefont {A.~W.~W.}\ \bibnamefont {Ludwig}},\
  }\href@noop {} {\bibfield  {journal} {\bibinfo  {journal} {New Journal of
  Physics}\ }\textbf {\bibinfo {volume} {12}},\ \bibinfo {pages} {065010}
  (\bibinfo {year} {2010})}\BibitemShut {NoStop}%
\bibitem [{\citenamefont {Fidkowski}\ and\ \citenamefont
  {Kitaev}(2010)}]{FidkowskiKitaev1}%
  \BibitemOpen
  \bibfield  {author} {\bibinfo {author} {\bibfnamefont {L.}~\bibnamefont
  {Fidkowski}}\ and\ \bibinfo {author} {\bibfnamefont {A.}~\bibnamefont
  {Kitaev}},\ }\href {\doibase 10.1103/PhysRevB.81.134509} {\bibfield
  {journal} {\bibinfo  {journal} {Phys. Rev. B}\ }\textbf {\bibinfo {volume}
  {81}},\ \bibinfo {pages} {134509} (\bibinfo {year} {2010})}\BibitemShut
  {NoStop}%
\bibitem [{\citenamefont {Fidkowski}\ and\ \citenamefont
  {Kitaev}(2011)}]{FidkowskiKitaev2}%
  \BibitemOpen
  \bibfield  {author} {\bibinfo {author} {\bibfnamefont {L.}~\bibnamefont
  {Fidkowski}}\ and\ \bibinfo {author} {\bibfnamefont {A.}~\bibnamefont
  {Kitaev}},\ }\href {\doibase 10.1103/PhysRevB.83.075103} {\bibfield
  {journal} {\bibinfo  {journal} {Phys. Rev. B}\ }\textbf {\bibinfo {volume}
  {83}},\ \bibinfo {pages} {075103} (\bibinfo {year} {2011})}\BibitemShut
  {NoStop}%
\bibitem [{\citenamefont {Beenakker}(1997)}]{BeenakkerTransport}%
  \BibitemOpen
  \bibfield  {author} {\bibinfo {author} {\bibfnamefont {C.~W.~J.}\
  \bibnamefont {Beenakker}},\ }\href {\doibase 10.1103/RevModPhys.69.731}
  {\bibfield  {journal} {\bibinfo  {journal} {Rev. Mod. Phys.}\ }\textbf
  {\bibinfo {volume} {69}},\ \bibinfo {pages} {731} (\bibinfo {year}
  {1997})}\BibitemShut {NoStop}%
\bibitem [{\citenamefont {Aleiner}\ \emph {et~al.}(2002)\citenamefont
  {Aleiner}, \citenamefont {Brouwer},\ and\ \citenamefont {Glazman}}]{Aleiner}%
  \BibitemOpen
  \bibfield  {author} {\bibinfo {author} {\bibfnamefont {I.}~\bibnamefont
  {Aleiner}}, \bibinfo {author} {\bibfnamefont {P.}~\bibnamefont {Brouwer}}, \
  and\ \bibinfo {author} {\bibfnamefont {L.}~\bibnamefont {Glazman}},\ }\href
  {\doibase http://dx.doi.org/10.1016/S0370-1573(01)00063-1} {\bibfield
  {journal} {\bibinfo  {journal} {Physics Reports}\ }\textbf {\bibinfo {volume}
  {358}},\ \bibinfo {pages} {309 } (\bibinfo {year} {2002})}\BibitemShut
  {NoStop}%
\bibitem [{Note1()}]{Note1}%
  \BibitemOpen
  \bibinfo {note} {We assume spinless fermions for simplicity; spin can be
  introduced trivially since we impose $\protect \mathcal {T}^2 = 1$
  symmetry.}\BibitemShut {Stop}%
\bibitem [{\citenamefont {Verbaarschot}(1994)}]{VerbaarschotRMT}%
  \BibitemOpen
  \bibfield  {author} {\bibinfo {author} {\bibfnamefont {J.}~\bibnamefont
  {Verbaarschot}},\ }\href {\doibase 10.1103/PhysRevLett.72.2531} {\bibfield
  {journal} {\bibinfo  {journal} {Phys. Rev. Lett.}\ }\textbf {\bibinfo
  {volume} {72}},\ \bibinfo {pages} {2531} (\bibinfo {year}
  {1994})}\BibitemShut {NoStop}%
\bibitem [{\citenamefont {Stephanov}\ \emph {et~al.}(2005)\citenamefont
  {Stephanov}, \citenamefont {Verbaarschot},\ and\ \citenamefont
  {Wettig}}]{StephanovRMT}%
  \BibitemOpen
  \bibfield  {author} {\bibinfo {author} {\bibfnamefont {M.~A.}\ \bibnamefont
  {Stephanov}}, \bibinfo {author} {\bibfnamefont {J.~J.~M.}\ \bibnamefont
  {Verbaarschot}}, \ and\ \bibinfo {author} {\bibfnamefont {T.}~\bibnamefont
  {Wettig}},\ }\href@noop {} {\  (\bibinfo {year} {2005})},\ \Eprint
  {http://arxiv.org/abs/arXiv:hep-ph/0509286} {arXiv:hep-ph/0509286}
  \BibitemShut {NoStop}%
\bibitem [{\citenamefont {Beenakker}(2015)}]{BeenakkerReview2}%
  \BibitemOpen
  \bibfield  {author} {\bibinfo {author} {\bibfnamefont {C.~W.~J.}\
  \bibnamefont {Beenakker}},\ }\href {\doibase 10.1103/RevModPhys.87.1037}
  {\bibfield  {journal} {\bibinfo  {journal} {Rev. Mod. Phys.}\ }\textbf
  {\bibinfo {volume} {87}},\ \bibinfo {pages} {1037} (\bibinfo {year}
  {2015})}\BibitemShut {NoStop}%
\bibitem [{\citenamefont {Guhr}\ \emph {et~al.}(1998)\citenamefont {Guhr},
  \citenamefont {Muller-Groeling},\ and\ \citenamefont
  {Weidenmuller}}]{GuhrRMT}%
  \BibitemOpen
  \bibfield  {author} {\bibinfo {author} {\bibfnamefont {T.}~\bibnamefont
  {Guhr}}, \bibinfo {author} {\bibfnamefont {A.}~\bibnamefont
  {Muller-Groeling}}, \ and\ \bibinfo {author} {\bibfnamefont {H.~A.}\
  \bibnamefont {Weidenmuller}},\ }\href {\doibase
  http://dx.doi.org/10.1016/S0370-1573(97)00088-4} {\bibfield  {journal}
  {\bibinfo  {journal} {Physics Reports}\ }\textbf {\bibinfo {volume} {299}},\
  \bibinfo {pages} {189 } (\bibinfo {year} {1998})}\BibitemShut {NoStop}%
\bibitem [{\citenamefont {Silverstein}(1985)}]{SilversteinWishart}%
  \BibitemOpen
  \bibfield  {author} {\bibinfo {author} {\bibfnamefont {J.~W.}\ \bibnamefont
  {Silverstein}},\ }\href {http://www.jstor.org/stable/2244186} {\bibfield
  {journal} {\bibinfo  {journal} {The Annals of Probability}\ }\textbf
  {\bibinfo {volume} {13}},\ \bibinfo {pages} {1364} (\bibinfo {year}
  {1985})}\BibitemShut {NoStop}%
\bibitem [{\citenamefont {Bai}\ and\ \citenamefont {Yin}(1993)}]{BaiWishart}%
  \BibitemOpen
  \bibfield  {author} {\bibinfo {author} {\bibfnamefont {Z.~D.}\ \bibnamefont
  {Bai}}\ and\ \bibinfo {author} {\bibfnamefont {Y.~Q.}\ \bibnamefont {Yin}},\
  }\href {\doibase 10.1214/aop/1176989118} {\bibfield  {journal} {\bibinfo
  {journal} {Ann. Probab.}\ }\textbf {\bibinfo {volume} {21}},\ \bibinfo
  {pages} {1275} (\bibinfo {year} {1993})}\BibitemShut {NoStop}%
\bibitem [{\citenamefont {Chiu}\ \emph {et~al.}(2015)\citenamefont {Chiu},
  \citenamefont {Pikulin},\ and\ \citenamefont {Franz}}]{PikulinFranz1}%
  \BibitemOpen
  \bibfield  {author} {\bibinfo {author} {\bibfnamefont {C.-K.}\ \bibnamefont
  {Chiu}}, \bibinfo {author} {\bibfnamefont {D.~I.}\ \bibnamefont {Pikulin}}, \
  and\ \bibinfo {author} {\bibfnamefont {M.}~\bibnamefont {Franz}},\ }\href
  {\doibase 10.1103/PhysRevB.92.241115} {\bibfield  {journal} {\bibinfo
  {journal} {Phys. Rev. B}\ }\textbf {\bibinfo {volume} {92}},\ \bibinfo
  {pages} {241115} (\bibinfo {year} {2015})}\BibitemShut {NoStop}%
\bibitem [{\citenamefont {Pikulin}\ \emph {et~al.}(2015)\citenamefont
  {Pikulin}, \citenamefont {Chiu}, \citenamefont {Zhu},\ and\ \citenamefont
  {Franz}}]{PikulinFranz2}%
  \BibitemOpen
  \bibfield  {author} {\bibinfo {author} {\bibfnamefont {D.~I.}\ \bibnamefont
  {Pikulin}}, \bibinfo {author} {\bibfnamefont {C.-K.}\ \bibnamefont {Chiu}},
  \bibinfo {author} {\bibfnamefont {X.}~\bibnamefont {Zhu}}, \ and\ \bibinfo
  {author} {\bibfnamefont {M.}~\bibnamefont {Franz}},\ }\href {\doibase
  10.1103/PhysRevB.92.075438} {\bibfield  {journal} {\bibinfo  {journal} {Phys.
  Rev. B}\ }\textbf {\bibinfo {volume} {92}},\ \bibinfo {pages} {075438}
  (\bibinfo {year} {2015})}\BibitemShut {NoStop}%
\bibitem [{\citenamefont {Gorin}(2002)}]{GorinIntegral}%
  \BibitemOpen
  \bibfield  {author} {\bibinfo {author} {\bibfnamefont {T.}~\bibnamefont
  {Gorin}},\ }\href {\doibase 10.1063/1.1471367} {\bibfield  {journal}
  {\bibinfo  {journal} {Journal of Mathematical Physics}\ }\textbf {\bibinfo
  {volume} {43}},\ \bibinfo {pages} {3342} (\bibinfo {year} {2002})},\ \Eprint
  {http://arxiv.org/abs/http://dx.doi.org/10.1063/1.1471367}
  {http://dx.doi.org/10.1063/1.1471367} \BibitemShut {NoStop}%
\bibitem [{\citenamefont {Prosen}\ \emph {et~al.}(2002)\citenamefont {Prosen},
  \citenamefont {Seligman},\ and\ \citenamefont
  {Weidenmuller}}]{ProsenIntegral}%
  \BibitemOpen
  \bibfield  {author} {\bibinfo {author} {\bibfnamefont {T.}~\bibnamefont
  {Prosen}}, \bibinfo {author} {\bibfnamefont {T.~H.}\ \bibnamefont
  {Seligman}}, \ and\ \bibinfo {author} {\bibfnamefont {H.~A.}\ \bibnamefont
  {Weidenmuller}},\ }\href {\doibase 10.1063/1.1506955} {\bibfield  {journal}
  {\bibinfo  {journal} {Journal of Mathematical Physics}\ }\textbf {\bibinfo
  {volume} {43}},\ \bibinfo {pages} {5135} (\bibinfo {year} {2002})},\ \Eprint
  {http://arxiv.org/abs/http://aip.scitation.org/doi/pdf/10.1063/1.1506955}
  {http://aip.scitation.org/doi/pdf/10.1063/1.1506955} \BibitemShut {NoStop}%
\bibitem [{Note2()}]{Note2}%
  \BibitemOpen
  \bibinfo {note} {Here connectedness refers to the way the $U$ tensors are
  contracted.}\BibitemShut {Stop}%
\bibitem [{\citenamefont {Bender}\ and\ \citenamefont
  {Canfield}(1978)}]{BENDERGraph}%
  \BibitemOpen
  \bibfield  {author} {\bibinfo {author} {\bibfnamefont {E.~A.}\ \bibnamefont
  {Bender}}\ and\ \bibinfo {author} {\bibfnamefont {E.}~\bibnamefont
  {Canfield}},\ }\href {\doibase
  http://dx.doi.org/10.1016/0097-3165(78)90059-6} {\bibfield  {journal}
  {\bibinfo  {journal} {Journal of Combinatorial Theory, Series A}\ }\textbf
  {\bibinfo {volume} {24}},\ \bibinfo {pages} {296 } (\bibinfo {year}
  {1978})}\BibitemShut {NoStop}%
\bibitem [{\citenamefont {Bollobas}(1982)}]{BollobasGraph}%
  \BibitemOpen
  \bibfield  {author} {\bibinfo {author} {\bibfnamefont {B.}~\bibnamefont
  {Bollobas}},\ }\href {\doibase 10.1112/jlms/s2-26.2.201} {\bibfield
  {journal} {\bibinfo  {journal} {Journal of the London Mathematical Society}\
  }\textbf {\bibinfo {volume} {s2-26}},\ \bibinfo {pages} {201} (\bibinfo
  {year} {1982})}\BibitemShut {NoStop}%
\end{thebibliography}%

\appendix*
\section{Corrections to Wick's theorem}

In this Appendix we discuss the asymptotically small corrections to Wick's theorem for the $J_{ijkl}$ couplings.  Our treatment is quite general and does not rely on our particular proposed realization.  We will take the best-case scenario for randomness, invoking Eq.~\eqref{Ndot_relation} and assuming completely disordered and independent zero-mode wavefunctions $\phi_i$ that obey Eq.~\eqref{phi_properties}.  In practice the $\phi_i$'s also suffer subdominant correlations and will not be truly Gaussian; these corrections can be studied using techniques described, e.g., in Refs.~\onlinecite{GorinIntegral, ProsenIntegral} but are neglected for simplicity.

Using Eq.~\eqref{Jijkl} we see that correlations among $m$ $J_{ijkl}$'s satisfy
\begin{eqnarray}
\langle \prod_{f = 1}^m J_{i_fj_fk_fl_f} \rangle &=& \left(\frac{4!}{2^4}\right)^m \prod_f U^{\rm as}_{a_fb_fc_fd_f}
  \nonumber \\
&\times& \langle \prod_f \phi_{i_fa_f} \phi_{j_fb_f} \phi_{k_fc_f} \phi_{l_fd_f}\rangle,
\label{GeneralFormofCorrection}
\end{eqnarray} 
where repeated indices are summed.  The assumption of i.i.d.~Gaussian wavefunction components $\phi_{i,I}$ allows us to simply apply Wick's theorem to evaluate the right-hand side---though this does not mean the $J_{ijkl}$ couplings are independent.  Each $U^{\rm as}_{a_fb_fc_fd_f}$ connects to four wavefunction elements $\phi_{i_f,a_f}$, etc., and each wavefunction element contracts with another.  The disorder average on the right side of Eq.~\eqref{GeneralFormofCorrection} thus pairs all the $a_f,b_f,c_f,d_f$ indices in some manner and also forces all of the $i_f, j_f, k_f, l_f$ indices to similarly pair together (otherwise the average vanishes trivially).  

Wick-theorem-obeying correlations among $J_{ijkl}$'s occur when all four indices of each $U^{\rm as}_{a_fb_fc_fd_f}$ pair with all four indices of another.  For a local interaction such cases yield
\begin{eqnarray}
  \langle \prod_{f = 1}^m J_{i_fj_fk_fl_f} \rangle_{\rm Wick} &\sim& N_{\rm dot}^{g-2m},
\end{eqnarray}   
where $g$ is the number of connected pieces in the diagram and $m$ must be even \footnote{Here connectedness refers to the way the $U$ tensors are contracted.}.  Note that the maximally disconnected pieces with $g = m/2$ minimize the decay with $N_{\rm dot}$ and reduce to $\langle J_{ijkl}^2\rangle^{m/2}$.  

Non-Wick correlations arise when more than two $J_{ijkl}$'s share indices.  Equation~\eqref{WickViolation} gives one example.  Another is
\begin{eqnarray}
  \langle J_{ijkl}J_{inkp}J_{mjol}J_{mnop} \rangle &\propto& 
  \nonumber \\
  && \!\!\!\!\!\!\!\!\!\!\!\!\!\!\!\!\!\!\!\!\!\!\!\!\!\!\!\!\!\!\!\!\!\!\!\!\!\!\!\!\!\!\!\!\!\!\!\!\!\!\!\!\!\!\!\!\! \!\!\!\!\!\!\!\!\!\!\!\!\!\!\!\!\!\!\!  \frac{1}{N_{\rm dot}^8}\sum_{abcdefgh} U^{\rm as}_{abcd}U^{\rm as}_{ebgd}U^{\rm as}_{afch} U^{\rm as}_{efgh} \sim N_{\rm dot}^{-7},
  \label{WickViolation2}
\end{eqnarray}
where we again used a local interaction.  By contrast, Eqs.~\eqref{WickViolation} and \eqref{WickViolation2} both vanish in the SYK model.  Such non-Wick contributions are generically suppressed by some power of $N_{\rm dot}$ compared to $\langle J_{ijkl}^2\rangle^{m/2}$.  Naively, this analysis suggests that the Wick contributions dominate the Feynman-diagram expansion of the model, and that hence we can use the `melon diagram' formalism \cite{KitaevTalks, Maldacena, Polchinski} that yields the large-$N$ SYK solution. 

However, at increasingly high order in the diagrammatic expansion, non-Wick correlations lead to a proliferation of new diagrams. Though each individual contribution is small in the above sense, the number of allowed graphs grows far faster than the suppression in $1/N_{\rm dot}$.  References \cite{BENDERGraph, BollobasGraph} discuss the asymptotic number of simple regular graphs, that is, graphs with each vertex connected to a fixed number of edges. In our case we are concerned with 4-regular graphs: the number scales as
\begin{equation}
  P(m) \sim \dfrac{(4m)!}{(2m)!(m)!C^m} \sim m^m
  \label{AsymptoticNumberOfGraphs}
\end{equation}
for some constant $C$.  Note that this result underestimates the number of graphs since it excludes the ones that have multiple edges that join the same vertices.  Thus while each graph is, at best,  suppressed exponentially in $m$ (i.e., by a factor $1/N_{\rm dot}^{{\rm const} \times m}$), the number of such graphs grows combinatorially, and it becomes very nontrivial to resum the Feynman diagrams for the two and four point functions.  Hence the melon and ladder diagram resummations may need to be amended.  

\end{document}